\documentclass[12pt]{article}
\usepackage{graphicx}
\usepackage{amssymb}
\usepackage{amsmath}
\usepackage{cite}

\newcommand\pubnumber{Cavendish-HEP-2011-20}
\newcommand\pubdate{\today}
\newcommand{\vect}[1]{\boldsymbol{#1}}

\def\support{\footnote{This report summarises work performed in collaboration with W. J. Stirling. JG is supported by the UK Science
and Technology Facilities Council.}}

\def\Title#1{\begin{center} {\Large #1 } \end{center}}
\def\Author#1{\begin{center}{ \sc #1} \end{center}}
\def\Address#1{\begin{center}{ \it #1} \end{center}}

\newcommand\pubblock{\rightline{\begin{tabular}{l} \pubnumber\\
         \pubdate  \end{tabular}}}
\newenvironment{Abstract}{\begin{quotation}  }{\end{quotation}}
\newenvironment{Presented}{\begin{quotation} \begin{center} 
             PRESENTED AT\end{center}\bigskip 
      \begin{center}\begin{large}}{\end{large}\end{center} \end{quotation}}

\def\beq{\begin{equation}}
\def\eeq#1{\label{#1}\end{equation}}
\def\eeqn{\end{equation}}

\def\beqa{\begin{eqnarray}}
\def\eeqa#1{\label{#1}\end{eqnarray}}
\def\eeqan{\end{eqnarray}}

\let\bar=\overbar

\def\Dslash{\not{\hbox{\kern-4pt $D$}}}
\def\dslash{\not{\hbox{\kern-2pt $\del$}}}

\def\msb{{\bar{\ssstyle M \kern -1pt S}}}

\begin{document}
\begin{titlepage}
\pubblock

\vfill
\Title{Double Parton Splitting Diagrams and Interference and Correlation Effects in Double Parton Scattering}
\vfill
\Author{Jonathan Gaunt\support}
\Address{Cavendish Laboratory, University of Cambridge,
J.J. Thomson Avenue, Cambridge CB3 0HE, U.K.}
\vfill
\begin{Abstract}
We discuss two topics in double parton scattering (DPS) theory that have been the subject of recent
research interest. First, the role of `double parton splitting' diagrams in DPS is discussed. We 
outline the `double PDF' description of DPS, which was introduced a number of years ago.
It is pointed out that under this framework, a certain structure is anticipated in the cross section
expression for a `double perturbative splitting' diagram, which in the framework is
 regarded as DPS. We show that although this structure does indeed appear in the `double perturbative 
splitting' diagrams, there is no natural reason to demarcate specifically this part of the graph as 
the DPS part, and indeed there is no natural part of these diagrams that can be regarded as DPS. There 
therefore appear to be some unsatisfactory features in the double PDF approach to describing
 DPS. The second issue we discuss is that of interference and correlated parton contributions to 
proton-proton DPS. We explain in simple terms why there can be such contributions to the DPS cross
section. The potential existence of such contributions was pointed out long ago by Mekhfi and more
recently by Diehl and Schafer, but has been largely ignored in phenomenological investigations of DPS.
\end{Abstract}
\vfill
\begin{Presented}
MPI$@$LHC 2010\\
Glasgow, UK, November 29 -- December 3, 2010
\end{Presented}
\vfill
\end{titlepage}
\def\thefootnote{\fnsymbol{footnote}}
\setcounter{footnote}{0}

\section{`Double Perturbative Splitting' Diagrams in \\ Double Parton Scattering} \label{sec:splittingdiags}

We define double parton scattering (DPS) as the process in which two pairs of partons participate in hard interactions in a 
single proton-proton (p-p) collision. DPS processes can constitute important backgrounds to Higgs and other interesting signals (see 
e.g. \cite{DelFabbro:1999tf}), and can themselves be considered as interesting signal processes, since they reveal information 
about parton pair correlations in the proton.

Making only the assumption that the hard processes A and B may be factorised, the cross section for proton-proton DPS 
may be written as follows:
\vspace{-0.3cm}
\begin{align} \label{DPSXsec1}
\sigma^D_{(A,B)} \propto &\sum_{i,j,k,l}\int \prod_{a=1}^{4}dx_a d^2\vect{b}
\hat{\sigma}_{ik \to A}(\hat{s} = x_1x_3s) \hat{\sigma}_{jl \to B}(\hat{s} = x_2x_4s) \\ \nonumber
&\times \Gamma_{ij}(x_1,x_2,\vect{b};Q_A^2,Q_B^2)\Gamma_{kl}(x_3,x_4,\vect{b};Q_A^2,Q_B^2) 
\end{align}

The cross section formula is somewhat similar to that used for single parton scattering (SPS), except that two 
parton-level cross sections $\hat{\sigma}$ appear, and the PDF factors are two-parton generalised PDFs $\Gamma$
(2pGPDs) rather than single PDFs. Note that in this formula the two 2pGPDs are integrated over a common parton 
pair transverse separation $\vect{b}$ -- the transverse separation must clearly be identical in both protons in 
order that two pairs of partons meet in two separate hard interactions A and B. The DPS cross section
cannot naturally be written in terms of PDFs individually integrated over their $\vect{b}$ arguments, as is the
case for the SPS cross section. 

In many extant studies of DPS, it is assumed that the 2pGPD can be approximately factorised into a product of a
longitudinal piece and a (typically flavour and scale independent) transverse piece:
\begin{equation} \label{2pGPDdecomp2dPDF}
\Gamma_{ij}(x_1,x_2,\vect{b};Q_A^2,Q_B^2) \simeq D_p^{ij}(x_1,x_2;Q_A^2,Q_B^2) F(\vect{b})
\end{equation}

Then, if one introduces the quantity $\sigma_{eff}$ via $\sigma_{eff} \equiv 1/[\int F(\vect{b})^2 d^2 \vect{b}]$, one 
finds that one may write $\sigma^D_{(A,B)}$ entirely in terms of the longitudinal piece and $\sigma_{eff}$:
\vspace{-0.4cm}
\begin{align} \label{dPDFXsec}
\sigma^{D}_{(A,B)} \propto& \dfrac{1}{\sigma_{eff}}\sum_{i,j,k,l}\int \prod_{a=1}^{4}dx_a D_p^{ij}(x_1,x_2;Q_A^2,Q_B^2) 
D_p^{kl}(x_3,x_4;Q_A^2,Q_B^2)
 \hat{\sigma}_{ik \to A} \hat{\sigma}_{jl \to B}
\end{align}

In \cite{Zinovev:1982be} a quantity $D_p^{ij}(x_1,x_2; Q^2)$ is introduced, and an evolution equation for this quantity is
given. We shall refer to the quantity and its evolution equation as the double PDF (dPDF) and the dDGLAP equation respectively. It is
asserted in \cite{Snigirev:2003cq} that the dPDF is equal to the factorised longitudinal part of the 2pGPD in the case in
which the two hard scales $Q_A^2$ and $Q_B^2$ are equal to a common value $Q^2$. 

\begin{figure}
\centering
\includegraphics[scale=0.6, trim = 0 1cm 0 1cm]{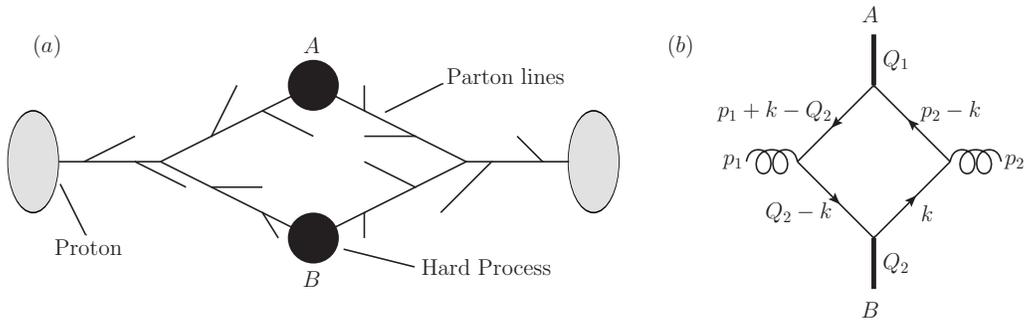}
\caption{\label{fig:dpsloops} (a) A diagram that apparently contributes to the leading order DPS cross section according to the
framework of \cite{Snigirev:2003cq}. 
(b) The `crossed box' graph. In this part of the figure, A and B are arbitrary single particle final states with 
$Q_1^2 = Q_2^2 = Q^2 > 0$.}
\end{figure}

The dDGLAP equation contains two types of terms on the right hand side -- `independent branching' terms corresponding to
emission of partons from a pre-existing pair, and `single parton feed' terms corresponding to the perturbative generation
of a pair from the splitting of a single parton. The single feed terms involve the leading twist single parton 
distributions as one might expect. Given this structure of the dDGLAP equation, with single feed terms included on the
right hand side, a prediction of the framework suggested in \cite{Snigirev:2003cq} for calculating the proton-proton DPS
cross section is that a part of the graph drawn in figure \ref{fig:dpsloops}(a) should be included in the DPS cross section. 
The part that should be included is proportional to $[\log(Q^2/\Lambda^2)]^n/\sigma_{eff}$ at the cross section level, 
where $\Lambda$ is some IR cutoff of order $\Lambda_{QCD}$, and $n$ is equal to the total number of QCD branchings in 
figure \ref{fig:dpsloops}(a) (including the two that only produce internal particles). This piece should be associated with
the region of transverse momentum integration for the graph in which the transverse momenta of the branchings on either
side of the `hard processes' in the graph are strongly ordered.

The question that then naturally arises is whether such a structure in fact exists in the cross section expression for the loop of
figure \ref{fig:dpsloops}(a). Starting from the conventional `Feynman rules' expression for the loop, it is not immediately obvious 
what the answer to this question is. Here we will focus on answering this question for the specific very simple `crossed box' loop shown in figure 
\ref{fig:dpsloops}(b), which is predicted by the framework of \cite{Snigirev:2003cq} to contain a piece proportional to $[\log(Q^2/\Lambda^2)]^2/\sigma_{eff}$.
The issues raised in the treatment of this example carry over directly to the more general loop of figure \ref{fig:dpsloops}(a).

We expect the $[\log(Q^2/\Lambda^2)]^2/\sigma_{eff}$ piece in figure \ref{fig:dpsloops}(b) to be predominantly contained in the
portion of the cross section integration in which the external transverse momenta, as well as the transverse momenta and virtualities
of the internal particles, are all small. This is actually the region around a certain Landau singularity in the loop integral known
as the double parton scattering singularity \cite{Nagy:2006xy}. In \cite{Gaunt:2011xd}, we obtained an analytic expression for the
part of an arbitrary loop containing a DPS singularity associated with the loop particles emerging from the initial state particles
being nearly on-shell and collinear, in the limit in which the external transverse momenta are small. Applied to the loop of figure
\ref{fig:dpsloops}(b) this reads (schematically, suppressing helicity and colour labels):
\begin{align} \label{DPSamp}
L_{DPS, \text{fig \ref{fig:dpsloops}(b)}} \propto \dfrac{1}{Q^2} & \int d^2 \vect{k}  \Phi_{g \to q\bar{q}} (x, \vect{k-Q_2})
 \Phi_{g \to \bar{q}q} (1-x, \vect{-k}) 
\\ \nonumber &\times 
\mathcal{M}_{q\bar{q} \to A} (\hat{s} = x(1-x)s) \mathcal{M}_{q\bar{q} \to B} (\hat{s} = x(1-x)s) + (q \leftrightarrow \bar{q})
\end{align}

In this formula, $x = p_2 \cdot Q_1 / p_1 \cdot p_2$, $s = (p_1+p_2)^2$, and $\vect{k}$ ($\vect{Q_2}$) is the component of 
$k$ ($Q_2$) transverse to the axis defined by the directions of the incoming particles.  $\Phi_{g \to q\bar{q}}(x,\vect{k})$ is the $\mathcal{O}(\alpha_S)$ light cone 
wavefunction (LCWF) to produce a $q\bar{q}$ pair from a $g$ \cite{Harindranath:1998pd}, with the quark having lightcone momentum 
fraction $x$ and transverse momentum $\vect{k}$ with respect to the parent gluon. It can be factored into a $\vect{k}$ and
$x$ dependent part, where the $\vect{k}$ dependent part is proportional to $\vect{\epsilon \cdot k}/\vect{k}^2$, $\vect{\epsilon}$
being the transverse part of the gluon polarisation vector. It is generally true that the $\vect{k}$ dependent part of the LCWF
 corresponding to any QCD splitting is proportional to $1/\vect{k}$.

Inserting \eqref{DPSamp} into the standard $2 \to 2$ cross section expression, and performing a number of changes of variable, we
arrive at the following expression for the DPS singular part of the $gg \to AB$ cross section:
\vspace{-0.3cm}
\begin{align} \label{ggDPSXsec2}
\sigma_{DPS, \text{fig \ref{fig:dpsloops}(b)}} \propto & \int \prod_{i=1}^{2}dx_id\bar{x}_i \hat{\sigma}_{q\bar{q} \to A}(\hat{s} = x_1\bar{x}_1s) 
\hat{\sigma}_{q\bar{q} \to B} (\hat{s} = x_2\bar{x}_2s)
\\ \nonumber
\times & \int \dfrac{d^2\vect{r}}{(2\pi)^2} 
\Gamma_{g \to q\bar{q}}(x_1,x_2, \vect{r}) \Gamma_{g \to \bar{q}q}(\bar{x}_1,\bar{x}_2, -\vect{r}) 
\end{align}
\vspace{-0.6cm}
\begin{align} \label{rspace2pGPD}
\quad \Gamma_{g \to q\bar{q}}(x_1,x_2, \vect{r}) \propto \dfrac{\alpha_S}{2\pi} \delta(1-x_1-x_2) T^{ij} (x_1,x_2)
 \int^{\vect{\tilde{k}}^2 < \mathcal{O}(Q^2)} d^2 \vect{\tilde{k}} \tfrac{[\vect{\tilde{k}}+\tfrac{1}{2}\vect{r}]^i [\vect{\tilde{k}}-\tfrac{1}{2}\vect{r}]^j}
{[\vect{\tilde{k}}+\tfrac{1}{2}\vect{r}]^2 [\vect{\tilde{k}}-\tfrac{1}{2}\vect{r}]^2}.
\end{align} 

$T^{ij} (x_1,x_2)$ contains a function of $x_1$ and $x_2$ that may be regarded as a `$1 \to 2$' splitting function, 
multiplied by a constant matrix in transverse space\footnote{Note that the cross section is really a sum of terms with 
different $T^{ij} (x_1,x_2)$ factors in the $g \to q\bar{q}$ 2pGPDs. This is associated with the fact that, from the 
point of view of the quarks, there is an unpolarised diagonal contribution to the process plus polarised and interference
contributions in colour, spin, and flavour space. See section \ref{sec:corrinter} for a discussion of correlation and interference
effects in DPS processes.}. $\vect{r}$ is equal to the transverse momentum imbalance of one of the 
quarks/antiquarks in the loop between amplitude and conjugate, and is the Fourier transform of the parton pair separation 
$\vect{b}$ in the $q\bar{q}$ pair emerging from either gluon. $\Gamma_{g \to q\bar{q}}(x_1,x_2, \vect{r})$ can therefore 
be thought of as the $\mathcal{O}(\alpha_S)$ transverse momentum-space 2pGPD to find a $q\bar{q}$ pair inside a gluon. Note that 
the expression here effectively coincides with that of \cite{Diehl:2011tt}, in which a cross section expression for the box of 
\ref{fig:dpsloops}(b) is obtained starting from a pure DPS view of the box.

Let us consider the part of the integral \eqref{ggDPSXsec2} that is associated with the magnitude of the 
imbalance $\vect{r}$ being smaller than some small cut-off $\Lambda$ that is of the order
of $\Lambda_{QCD}$. The contribution to the cross section from this portion contains a $\log^2(Q^2/\Lambda^2)$ factor 
multiplied by $\Lambda^2$ (which can be thought of as an effective `$1/\sigma_{eff}$' factor for this contribution). 
The majority of this contribution comes from the region in which the transverse momenta and virtualities of the 
quarks and antiquarks in the $gg \to AB$ loop are much smaller in magnitude than $\sqrt{Q^2}$ (i.e. the region in 
which the assumptions used to derive \eqref{DPSamp} apply), which is a necessary feature of a contribution to 
be able to regard it as a DPS-type contribution. By making a specific choice of $\Lambda$ (let us call this $\Lambda_S$), 
one could obtain an expression which is exactly in accord with the expectations of \cite{Snigirev:2003cq} -- that is, a product of 
two large DGLAP logarithms multiplied by the same $1/\sigma_{eff}$ factor that appears in diagrams in which the parton
pair from neither proton has arisen as a result of one parton perturbatively splitting into two.
The $1/\sigma_{eff}$ factor for these diagrams presumably has a natural value of the order of $1/R_p^2$ that is set
by the nonperturbative dynamics ($R_p$ = proton radius).

The fact that we have to make a somewhat arbitrary choice for $\Lambda$ in order to arrive at the result anticipated
by the framework of \cite{Snigirev:2003cq} is concerning. There is nothing in the calculation of figure \ref{fig:dpsloops}(b) 
to indicate that we should take the region of it with $|\vect{r}| < \Lambda_S$ as the `DPS part' -- the scale $\Lambda_S$ does not 
naturally appear at any stage of the calculation. There is no more justification for taking the part of the box with
$|\vect{r}| < \Lambda_S$ to be the DPS part than there is for, say, taking the piece with $|\vect{r}| < 2\Lambda_S$,
or that with $|\vect{r}| < \Lambda_S/2$, to be the DPS part.

There therefore appear to be some unsatisfactory features of the framework of \cite{Snigirev:2003cq} with regards 
to its treatment of the box in figure \ref{fig:dpsloops}(b). It is clear that these issues will also be encountered
for the case of the arbitrary `double perturbative splitting' graph in figure \ref{fig:dpsloops}(a). One obtains
a result that is consistent with the framework of \cite{Snigirev:2003cq} if one demarcates the portion of the cross
section integral in which the transverse loop momentum imbalance between amplitude and conjugate is less than
$\Lambda_S$ as DPS, but there is no natural reason to do this. There is no distinct piece of figure 
\ref{fig:dpsloops}(a) that contains a natural scale of order $\Lambda_{QCD}$ and is associated with the transverse
momenta inside the loop being strongly ordered on either side of the diagram -- most of the contribution to the
cross section expression for this graph comes from the region of integration in which the transverse momenta of 
particles inside the loop are of $\mathcal{O}(\sqrt{Q^2})$ (except perhaps at low $x$ values for the `hard
subprocesses' in the graph \cite{Ryskin:2011kk}). It is therefore perhaps the case that we should not 
regard any of this graph as DPS. Treating the graph in this way has the 
advantage that we do not perform any double counting between DPS and SPS -- the graph of figure \ref{fig:dpsloops}(a)
is in principle also included in the SPS $pp \to AB$ cross section (albeit as a very high order correction that will 
not be included in practical low order calculations, if the number of QCD emissions from inside the loop of the 
graph is large). 

One can gain some insight into the source of the problems in the framework of \cite{Snigirev:2003cq} by looking at
the $\vect{b}$-space 2pGPD corresponding to \eqref{rspace2pGPD}. This comes out as being proportional to $1/\vect{b}^2$
 -- this behaviour (which was first spotted in \cite{Diehl:2011tt}) can be traced to the fact that the $g \to q\bar{q}$ 
LCWF in $\vect{b}$ space (like any LCWF corresponding to a QCD perturbative splitting) is proportional to $1/\vect{b}$, 
and $\Gamma(\vect{b}) \sim \Phi(\vect{b})^2$. Note that this behaviour is very different from the behaviour of all
2pGPDs that is anticipated by the dPDF framework (i.e. smooth function of size $R_p$). There is no natural feature 
in the product of two `perturbative splitting' 2pGPDs that is of size $R_p$ and can be naturally identified as DPS. 
A key error then in the formulation of the dPDF framework seems to be the assumption that all 2pGPDs can be 
approximately factorised into dPDFs and smooth transverse functions of size $R_p$. A sound theoretical framework 
for describing proton-proton DPS needs to carefully take account of the different $\vect{b}$ dependence of pairs of 
partons emerging from perturbative splittings, whilst simultaneously avoiding double counting between SPS and DPS.
For recent proposals from other authors as to how proton-proton DPS should be described theoretically, including
further discussion of the `double parton splitting graphs', see \cite{Ryskin:2011kk, Blok:2011bu}.\footnote{It is 
worth pointing out in passing that there is perhaps a double scattering process which does 
directly involve the dPDF of the proton. This is the contribution to proton-heavy nucleus DPS associated with partons
from two separate nucleons interacting with two partons from the proton. The reason why this probes the dPDF, and 
p-p DPS does not, is that in this case the `probe' parton pair coming from the nucleus has a (roughly) flat
distribution in $\vect{b}$, whereas in p-p DPS some parton pairs (i.e. perturbatively generated ones) do not. For more
details and a discussion of how the two-nucleon contribution to proton-heavy nucleus DPS might be extracted experimentally,
see \cite{Strikman:2001gz}.}

\vspace{-0.4cm}

\section{Interference and Correlation Effects in DPS} \label{sec:corrinter}

It was pointed out long ago by Mekhfi \cite{Mekhfi:1985dv} that there can be contributions to
the p-p DPS cross section associated with spin and colour correlations and interference 
effects in spin and colour space, even when the colliding protons are unpolarised. The issue was taken
up again recently by Diehl and Schafer \cite{Diehl:2011tt} who demonstrated that the correlation
and interference contributions may be sizeable, and made the observation that interference effects
in flavour space can also contribute to unpolarised p-p DPS. In this section we explain in 
simple terms why there can be interference and correlated parton contributions to the unpolarised p-p 
DPS cross section, where there are no such contributions to the corresponding SPS cross section. 
We hope that this explanation may be of aid to those less familiar with the subject, and refer the reader 
to \cite{Mekhfi:1985dv, Diehl:2011tt} for more details.

\begin{figure}
\centering
\includegraphics[scale=0.4, trim = 0 1cm 0 0.3cm]{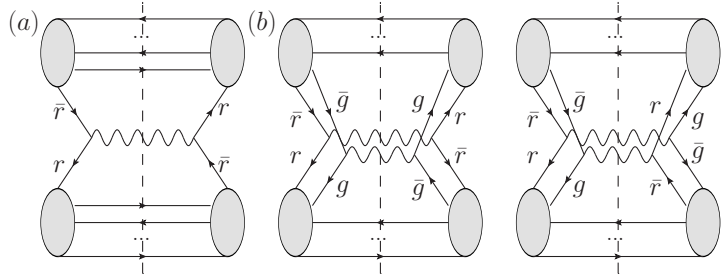}
\caption{\label{fig:DPSInterference}(a) Leading twist diagram for single Drell-Yan (as an example process) 
in proton-proton collisions. (b) A diagonal in colour (left) and colour interference (right) diagram 
contributing to the double Drell-Yan DPS process.}
\end{figure}

We recall that the cross section for leading twist single parton scattering processes is calculated from 
`cut diagrams' with the structure of figure \ref{fig:DPSInterference}(a). For definiteness we have taken
the SPS process to be Drell-Yan in the figure, but the details of the final state are not important for 
our discussion. Now, if we consider the parton `returning' to (say) the bottom proton on the right hand side 
of the diagram, then we see that it must have exactly the same flavour and colour as it `left' with on the 
left hand side. This must be the case otherwise it cannot `reform' the original proton when it combines with 
the spectators on the right hand side. So there can be no flavour and colour interference contributions to
p-p SPS. When the colliding protons are unpolarised, symmetry forbids any contribution to the SPS
cross section associated with helicity or transversity polarisation effects. For similar reasons, there 
cannot be any contribution to the SPS cross section associated with the analogous effects in colour space.
The only PDFs that contribute to the unpolarised SPS cross section are therefore the unpolarised diagonal 
colour-summed PDFs.

The cross section for DPS processes is calculated from cut diagrams with the structure of figure 
\ref{fig:DPSInterference}(b) in which two partons `leave' each proton on the left, interact, and then
`return' on the right. In this case, the fact that the proton must be reformed at the end only imposes 
constraints on the `sum' of the two partons' quantum numbers. For any DPS process, the possibility arises
for interference diagrams to contribute in which one or more of the discrete quantum numbers get swapped 
between the partons before they return to the proton -- provided that a swap in the opposite direction 
happens in the other proton (an example of a colour interference diagram that contributes to double Drell-Yan 
is the right hand graph of figure \ref{fig:DPSInterference}(b)). 

There are also contributions to the DPS cross section associated with polarised 2pGPDs that can
be nonzero even when the colliding protons are unpolarised. The reason for this is that, with two partons
participating from each proton in DPS, there can be effects relating to the correlations in spin between
the active partons themselves. So, if for example the chance to find a pair of quarks in the proton with 
their helicities aligned differs from that to find a pair of quarks with opposing helicities, then 
$\Delta q_1 \Delta q_2 \equiv q_1\hspace{-1mm}\uparrow q_2\hspace{-1mm}\uparrow + q_1\hspace{-1mm}\downarrow 
q_2\hspace{-1mm}\downarrow - q_1\hspace{-1mm}\downarrow q_2\hspace{-1mm}\uparrow - q_1\hspace{-1mm}\uparrow 
q_2\hspace{-1mm}\downarrow \neq 0$. In a similar way there will be contributions to p-p DPS associated 
with colour correlations.

Despite it being pointed out long ago that p-p DPS may be affected by interference and correlated parton 
effects, such effects are generally not considered in phenomenological analyses of this process. More detailed studies
need to be performed on these effects, including an examination of the effect of evolution on them (to what extent does 
evolution `wash out' correlations?) and it would be desirable to have more refined estimates 
of the size of interference and correlated parton contributions than were performed in \cite{Diehl:2011tt}.
One would require low-scale inputs for the interference and correlated parton two-parton distributions in order to make 
such estimates (along with the requisite evolution framework), very approximate forms for which could perhaps be 
extracted from proton models. Such forms would of course not be reliable at low $x$ owing to the fact that one cannot
fit parton densities at low $x$ even at low $Q^2$ without including a number of `nonperturbative' gluons and sea 
quarks \cite{Gluck:2007ck}, and proton models typically only include the lowest few Fock states. An alternative approach for obtaining 
`first guess' inputs for some of the distributions via single-parton GPDs is given in \cite{Diehl:2011tt}.

\providecommand{\href}[2]{#2}\begingroup\raggedright\endgroup

\end{document}